\documentclass[a4paper,10pt,conference]{IEEEtran}
\IEEEoverridecommandlockouts
\usepackage{cite}
\usepackage{amsmath,amssymb,amsfonts}
\usepackage{algorithmic}
\usepackage{graphicx}
\usepackage{textcomp}
\usepackage{xcolor}
\def\BibTeX{{\rm B\kern-.05em{\sc i\kern-.025em b}\kern-.08em
    T\kern-.1667em\lower.7ex\hbox{E}\kern-.125emX}}
\begin{document}
\title{A Privacy-Preserving Image Retrieval Scheme with a Mixture of Plain and EtC Images}

\author{\IEEEauthorblockN{Kenta Iida and Hitoshi Kiya}
\IEEEauthorblockA{Tokyo Metropolitan University, Tokyo, Japan\\
E-mail: iida-kenta2@ed.tmu.ac.jp, kiya@tmu.ac.jp}}

\maketitle

\begin{abstract}
In this paper, we propose a novel content-based image-retrieval scheme that allows us to use a mixture of plain images and compressible encrypted ones called ``encryption-then-compression (EtC) images." 
In the proposed scheme, extended SIMPLE descriptors are extracted from EtC images as well as from plain ones, so the mixed use of plain and encrypted images is available for image retrieval.
%In the proposed scheme, after applying negative-positive transformation into both  EtC and  plain images, extended SIMPLE descriptors are extracted from the transformed images.
%The use of descriptors extracted from transformed images allows us to retrieve images under the mixture of plain and encrypted images.
%The use of EtC images enables us not only to protect the visual information of images but also to apply image compression such as JPEG compression to encrypted images.
%In the proposed scheme, extended SIMPLE descriptors are extracted from EtC images as well as from plain ones, after applying negative-positive transformation into all query images. 
%The use of this transformation allows us to retrieve images under the mixed use of plain and encrypted images.
In an experiment, the proposed scheme was demonstrated to have almost the same retrieval performance as that for plain images, even with a mixture of plain and encrypted images.
\end{abstract}

%%%%%%%%%%%%%%%%%%%%%%%%%%%%%%%%%%%%%%%%%%%%%%%%%%
\section{Introduction}
\noindent With the growth of cloud environments, a large number of images have been uploaded to cloud storage and photo sharing services.
Most of these images include sensitive information, such as personal data and copyrights.
However, there is the possibility of data leakage and unauthorized use by service providers because they are not trusted in general.
Therefore, various privacy-preserving image identification \cite{EID} and retrieval \cite{PPIR, PPIR2,PPIR3,EIR1,EIR2,EIR3, EIR4,EIR5,EIR6} and processing schemes \cite{PPP1,PPP2} have been proposed.
%Therefore, various privacy-preserving image identification and retrieval and processing schemes have been proposed\cite{PPIR, PPIR2,PPIR3,EIR1,EIR2,EIR3, EIR4,EIR5,EIR6, EID,PPP1,PPP2,id}.

%For privacy-preserving image-retrieval on cloud services, it is required that encrypted images can be compressed because images are generally uploaded and stored in a compressed form to reduce the amount of data.
On the other hand, it is required for image identification and retrieval on cloud services that the schemes have robustness against the image compression\cite{PPIR, PPIR2,PPIR3,EID,id}.
This is because images are generally uploaded and stored in a compressed form to reduce the amount of data.
Thus, the use of compressible encrypted images is required for privacy-preserving image retrieval.
For above reasons, privacy-preserving image-retrieval methods should satisfy three requirements: 1) protecting the visual information of plain images, 2) achieving a high retrieval performance without decryption, and 3) using compressible encrypted images.
Requirement 1) comprises two requirements: 1-a) protecting images stored in databases of cloud service providers and 1-b) protecting query images uploaded by users.
Requirement 1-a) has to be always satisfied for privacy-preserving image-retrieval.
In contrast, if users do not care about the unauthorized use of query images, requirement 1-b) is not needed.
The proposed method enables users to choose whether requirement 1-b) is needed, under requirements 1-a), 2), and 3).

Encryption-then-compression (EtC) systems have been developed \cite{EtC} as systems that satisfy both requirements 1) and 3), but requirement 2) is not considered.
In this paper, we focus on a block scrambling-based image encryption method that was proposed for EtC systems, where images encrypted by the method are referred to as ``EtC images."

To make the retrieval scheme user-friendly, a content-based image-retrieval scheme for EtC images was proposed \cite{PPIR2}.
In this scheme, the use of EtC images with extended SIMPLE descriptors (E-SIMPLEs) achieves a high retrieval performance.
However,  EtC images have to be generated without applying negative-positive transformation, so that robustness of EtC images against ciphertext-only attacks degrade for image-retrieval.

%To satisfy all requirements, a content-based image-retrieval scheme for EtC images was proposed \cite{PPIR1}.
%In this scheme, the use of EtC images with extended SIMPLE descriptors (E-SIMPLEs) achieves a high retrieval performance.
%However, using plain images as query images has never been considered.
%Therefore, users cannot decide whether requirement 1-b) is needed.
%To make the retrieval scheme user-friendly, we extend the privacy preserving image-retrieval scheme in \cite{PPIR1} in this paper.
%The proposed method enables users to choose whether requirement 1-b) is needed, under requirements 1-a), 2), and 3).

Due to such a situation, the proposed scheme enables users to choose whether requirement 1-b) is needed, under requirements 1-a), 2), and 3), while EtC images are generated with applying negative-positive transformation.
In the proposed scheme, encryption-then-compression (EtC) images are used as compressible images \cite{EtC}. 
For image-retrieval, extended SIMPLE descriptors \cite{PPIR,PPIR2,PPIR3} are used for avoiding the influence of image encryption.
%For image-retrieval, after applying negative-positive transformation into both EtC and  plain images, extended SIMPLE descriptors \cite{PPIR} are extracted from the transformed images.
The proposed scheme was demonstrated to have the same retrieval performance as that of using plain images,  even when using a mixture of encrypted and plain images.

%%%%%%%%%%%%%%%%%%%%%%%%%%%%%%%%%%%%%%%%%%%%%%%%%%
\section{Related work} \label{sec:relatedWork}
\begin{figure}[t!]
\begin{center}
\begin{tabular}{c}
\begin{minipage}{0.5\hsize}
  \begin{center}
   \includegraphics[width=30mm]{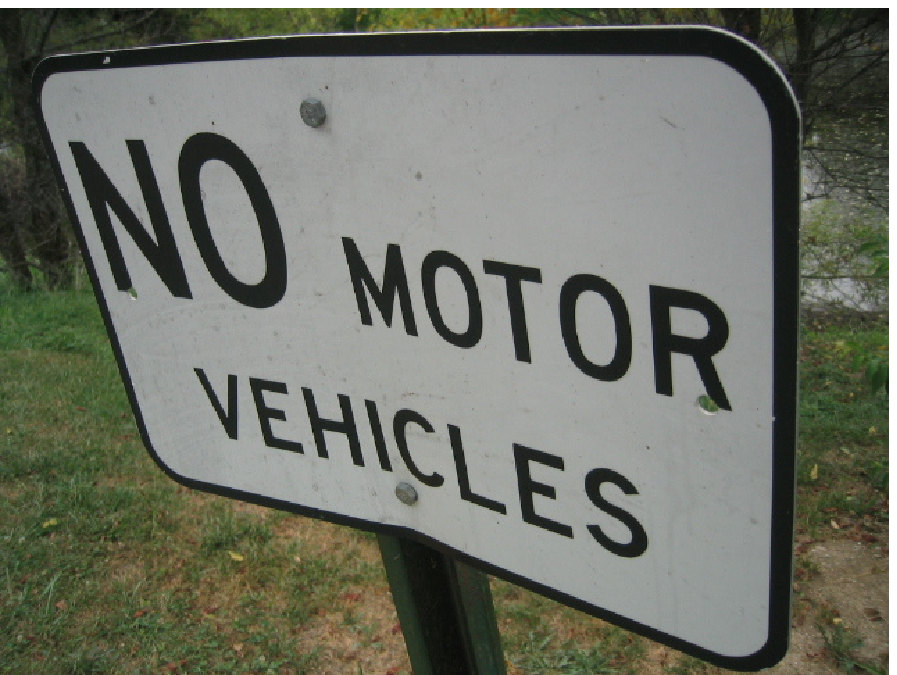}
     \hspace{20mm}(a) Plain image
  \end{center}
 \end{minipage}
\begin{minipage}{0.5\hsize}
  \begin{center}
   \includegraphics[width=30mm]{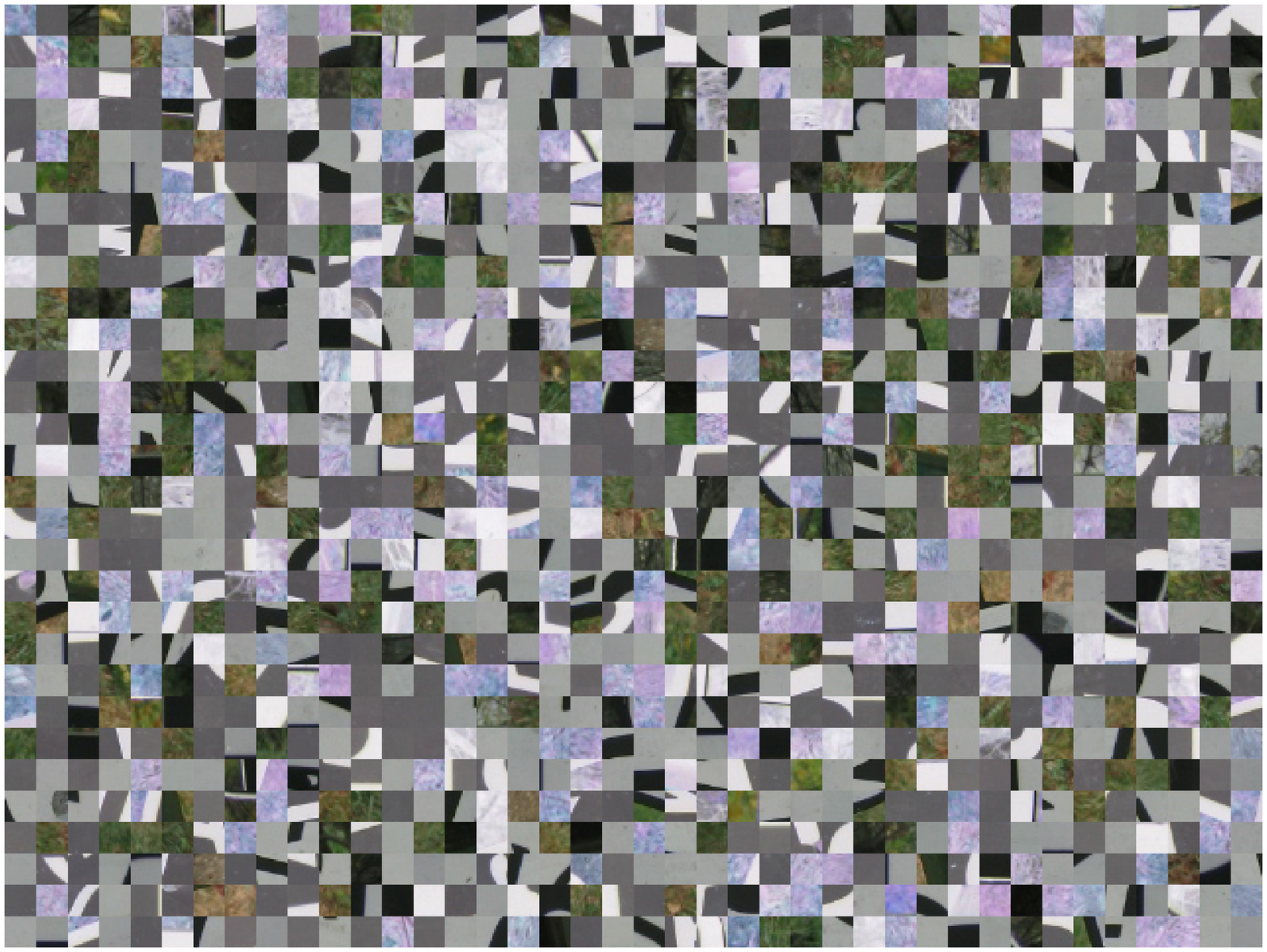}
     \hspace{20mm}(b) EtC image
  \end{center}
 \end{minipage}
\end{tabular}
\caption{Example of plain image and encrypted one\label{fig:encgen}}
 \end{center}
\end{figure} 
\begin{figure*}[t]
\centering
\begin{tabular}{cc}
\includegraphics[width=85mm]{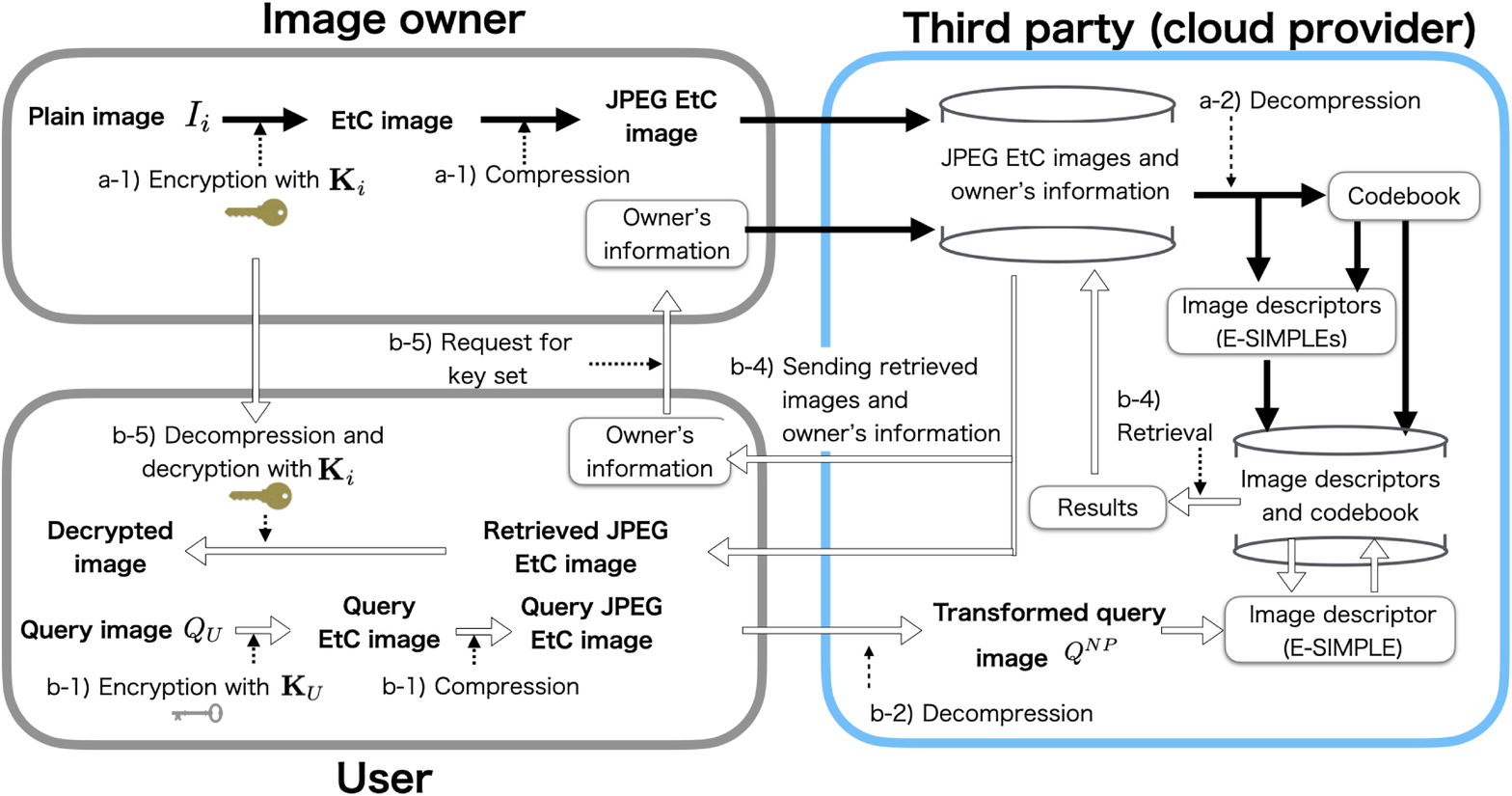} & \includegraphics[width=85mm]{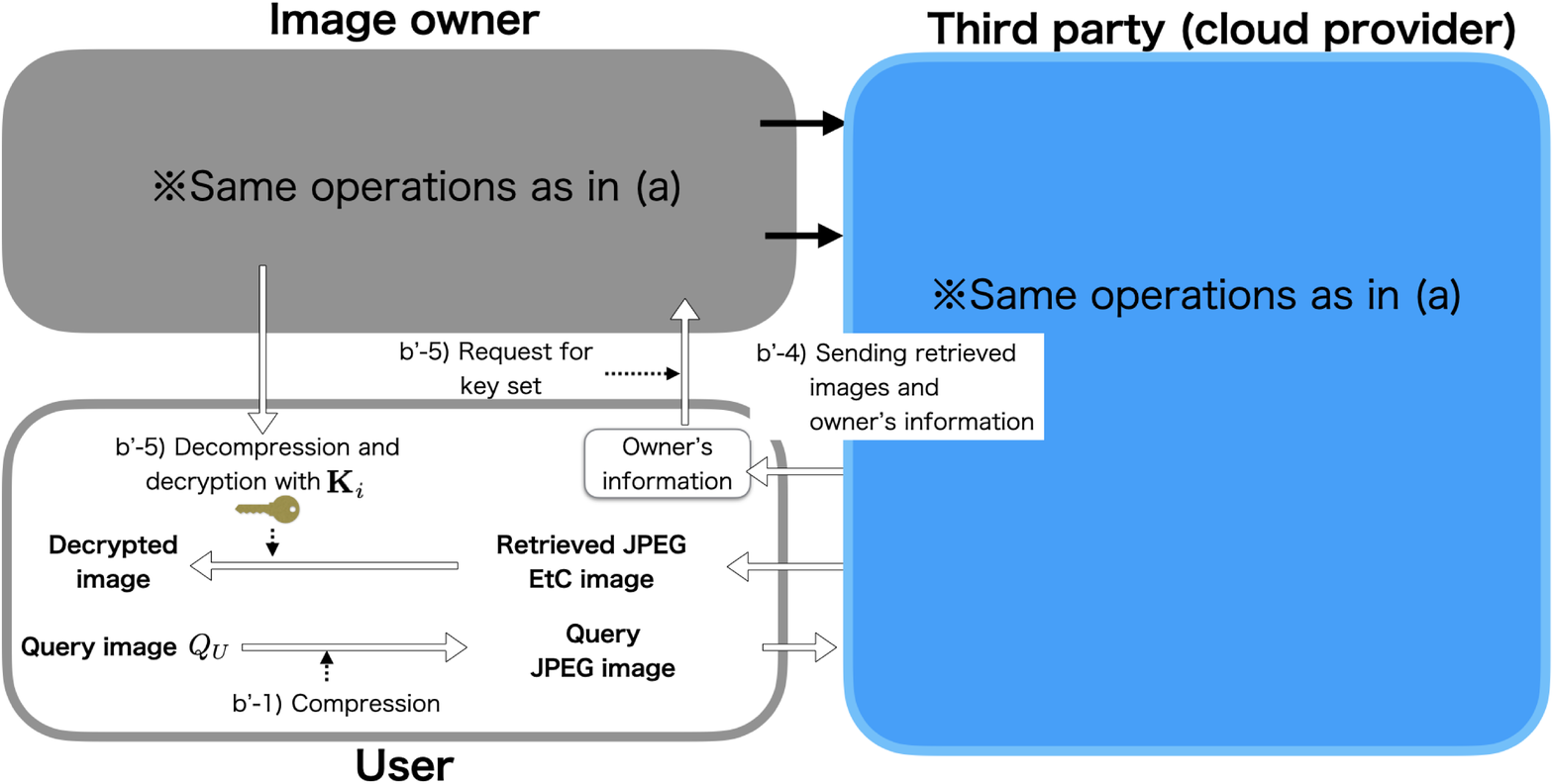}\\
(a) System model with protected query images & (b) System model with plain query images\\
\end{tabular}
\caption{System models, where image descriptors correspond to extended SIMPLE descriptors\label{fig:scenario}}
\end{figure*}
%%%%%%%%%%%%%%%%%%%%%%%%%%%%%%%%%%%%%%%%%%%%%%%%%%
\subsection{EtC image\label{sec:EtC}}
\noindent EtC images are images encrypted by using a block-wise encryption method proposed for encryption-then-compression (EtC) systems (see Fig. \ref{fig:encgen})\cite{EtC}.
EtC images not only have almost the same compression performance as that of plain images but also enough robustness against various ciphertext-only attacks. 
%EtC images can be compressed by using numerous compression methods such as JPEG. 
%In addition, they are robust enough against various ciphertext-only attacks including jigsaw puzzle solver attacks.

%To generate an EtC image with as a key set $\mathbf{K_i}=[K_i(1),K_i(2),K_i(3)]$, a plain image with a size of $X \times Y$ is divided into non-overlapping 16$\times$16 blocks, and $M = \lfloor \frac{X}{16} \rfloor \times \lfloor \frac{Y}{16} \rfloor$ blocks are obtained at first.
To generate an EtC image with a key set $\mathbf{K_i}=[K_i(1),K_i(2),K_i(3)]$,  a plain image with a size of $X \times Y$  into non-overlapping 16$\times$16 blocks at first.
Next, these divided blocks are randomly permuted by using a random integer secret key ${ K}_1(i)$.  
After that, each divided block is rotated and inverted by using a random integer  secret key ${ K}_2(i)$.
At last, negative-positive transformation is applied to every block by using a random binary integer generated by secret key $K_i(3)$.
A transformed pixel value in the $j$th block $B_j$, $p'$ is computed by
\begin{equation}
\label{eq:np}
\begin{cases}
p'=p,\ r(j)=0,\\
p' = 255 - p,\ r(j)=1,
\end{cases}
\end{equation}
where $r(j)$ is a random binary integer generated by $K_i(3)$ under the probability $P(r(j)) = 0.5$, and $p$ is the pixel value of a plain image with 8 bpp.

%In this paper, images encrypted by using these steps are referred to as ``EtC images.'' 
%$K_i(1)$, $K_i(2)$, and $K_i(3)$ are stored as a key set, $\mathbf{K_i}=[K_i(1),K_i(2),K_i(3)]$.

In the conventional privacy-preserving image retrieval scheme with a mixture use of plain and EtC images, negative-positive transformation can not be applied for keeping retrieval performances, so that robustness against the attacks degrades.
In contrast, the proposed scheme enables to apply negative-positive transformation, while keeping retrieval performances.

%%%%%%%%%%%%%%%%%%%%%%%%%%%%%%%%%%%%%%%%%%%%%%%%%%
\section{Proposed Scheme}
%%%%%%%%%%%%%%%%%%%%%%%%%%%%%%%%%%%%%%%%%%%%%%%%%%
\subsection{Extended SIMPLE descriptors}
\noindent In the proposed scheme, extended SIMPLE descriptors (E-SIMPLEs) are used as image descriptors for image retrieval\cite{PPIR,PPIR2,PPIR3} because of high retrieval performances for EtC images as well as one for plain images.

In the generation process of E-SIMPLEs, each image is divided into non-overlapping $16\times16$-blocks at first.
All $16\times16$-blocks are selected as patches, and then scalable color descriptor (SCD) is extracted as a patch descriptor from each patch.
By using $k$-means clustering, all extracted patch descriptors are classified into $M$ classes, and the set of $M$ centroid vectors are obtained as a codebook with a size of $M$.%After that, a codebook with a size of $M$ is generated from  the patch descriptors extracted from patches of all images by using k-means clustering.
After that, a histogram vector of each image is calculated by using the codebook and patch descriptors extracted from the image, and then the histogram vectors are weighted to obtain extended SIMPLE descriptors.
In the weighting process, when there are $N$ histogram vectors,  the $m$th component of the $n$th vectors $v_n(m)$, $0\leq m < M$, $0 \leq n < N$, is calculated as below in this paper.
\begin{equation}
v_n(m)=(1+log(tf_{(m,n)}))\times log\frac{N}{df_{(m)}},
\end{equation}
where $tf(m,n)$ represents the frequency of the $m$th component in the $n$th histogram vector, and $df(m)$ denotes the number of histogram vectors having non zero values in the $m$th component. 
After that, $l_2$ normalization is applied to every weighted histogram vector.

%\begin{itemize}
%\item[a)] Divide each image into non-overlapping $16\times16$-blocks, and use all $16\times16$-blocks as patches.
%%, where $16\times16$ corresponds to the block size of EtC images.
%\item[b)] Extract scalable color descriptor as a patch descriptor from each patch.
%\item[c)] Generate a codebook with a size of $M$ from the patch descriptors extracted from all images by using k-means clustering.
%\item[d)] Calculate a histogram vector of each image by using the codebook and patch descriptors extracted from the image.
%\item[e)] Obtain extended SIMPLE descriptors by weighting the histogram vectors.
%\end{itemize}. 
%For more details, kindly refer to \cite{PPIR}.
%In step e), all histogram vectors are weighted in order to obtain extended SIMPLE descriptors. 
%When $N$ histogram vectors are generated in step d), the $m$th component of the $n$th vectors $v_n(m)$, $0\leq m < M$, $0 \leq n < N$, is calculated as below in this paper.
%\begin{equation}
%v_n(m)=(1+log(tf_{(m,n)}))\times log\frac{N}{df_{(m)}},
%\end{equation}
%where $tf(m,n)$ represents the frequency of the $m$th visual word in the $n$th histogram vector, and $df(m)$ denotes the number of histogram vectors containing the $m$th visual word in the $N$ histogram vectors. 
%After that, $l_2$ normalization is applied to every weighted histogram vector.

The use of 16$\times$16-block sampling and SCD enables to avoid the influences of block scrambling and block rotation and inversion in principle.
However, due to the influence of negative-positive transformation, retrieval performances with E-SIMPLEs under the mixed use will degrade, as shown later.
%Although E-SIMPLEs enable to avoid the influences of block scrambling, block rotation and block rotation in principle, the influence of negative-positive transform is not guaranteed.
%Thus, this influence is considered for image retrieval under the mixed use in the proposed scheme. 

%%%%%%%%%%%%%%%%%%%%%%%%%%%%%%%%%%%%%%%%%%%%%%%%%%
\subsection{System Model}
%\begin{figure}[t!]
%\centering
%\begin{tabular}{c}
%\includegraphics[width=90mm]{./Sys1.eps}\\
%(a) System model with protected query images\\
%\includegraphics[width=90mm]{./Sys2.eps}\\
%(b) System model with plain query images\\
%\end{tabular}
%\caption{System models, where image descriptors correspond to extended SIMPLE descriptors\label{fig:scenario}}
%\end{figure}
\noindent Two system models used for the proposed method are shown in Fig. \ref{fig:scenario}, where the difference between the two models is whether query images are encrypted or not.
In the models, there are three roles: image owner, third party, and user, where the third party is not trusted.
The third party has image owners' information and EtC images uploaded from image owners and users, and moreover, it knows the encryption algorithm for generating EtC images.
The proposed method is designed not only to achieve a high retrieval performance but also to protect the visual information of plain images against attacks by the third party because the third party might try to restore the visual information of plain images from the EtC ones.
Here, the details of each operation performed in these models are summarized as below.
\subsection*{1) Process for generating image descriptors}
In both models, the following operations are conducted to generate image descriptors from images stored in the database of a third party.
\begin{enumerate}
\item[a-1)] 
An image owner generates an EtC image from plain image $I_i$ with secret key set ${\bf K}_i$, and the EtC image is then compressed with JPEG compression/a lossless-compression method.
The compressed EtC images are uploaded to a third party.

\item[a-2)]  
The third party generates a codebook from the uploaded EtC images after decompression, and image descriptors are then calculated from EtC images by using the codebook.
After that, the codebook and the image descriptors are stored in a database.

\end{enumerate}

JPEG compression is a lossy-compression method, so retrieved images contain some distortions due to the influence of image compression. By applying a lossless compression method to EtC images, users can restore original plain text images from received encrypted images without any degradation in image quality.

\subsection*{2) Retrieval process with protected query images}
\begin{enumerate}
\item[b-1)]  
\noindent A user sends query image ${ Q}_U^e$ encrypted by using key set ${\bf K}_U$ to a third party, where ${\bf K}_U$ can be prepared by the user.

\item[b-2)] 
The third party applies the block-wise negative-positive transformation to ${ Q}_U^e$. In this step, a transformed pixel value in the $j$th block $B_j$ , $p'$ is computed by

\begin{equation}
\label{eq:np}
\begin{cases}
p'=p,\ j=\textrm { 0, 2, 4,} \cdots\\
p' = 255 - p,\  j=\textrm { 1, 3, 5,} \cdots
\end{cases}
\end{equation} 
where $p$ is the pixel value of ${ Q}_U^e$ with 8 bpp, and the image with transformed pixel values is referred to as $Q^{NP}$. 

\item[b-3)] 
The third party calculates an image descriptor from ${ Q}^{NP}$ by using the stored codebook and the stored image descriptors.

\item[b-4)] 
The third party retrieves EtC images in the database similar to the query image by using the image descriptor in the encrypted domain. The retrieved images and the owner's information are returned to the user.

\item[b-5)] 
The user requests the image owner to send key sets for decrypting the EtC images received from the third party.
\end{enumerate}
In this framework, the third party not only has no visual information of images but also no secret keys. As demonstrated later, operation b-2) allows us to use a mixture of plain and encrypted images.

\subsection*{3) Retrieval process with plain query images}
\noindent If a user wants to use plain query images, the following steps are carried out.

\begin{enumerate}
\item[b'-1)]  
A user sends a query image $Q_U$ without any encryption to a third party.

\item[b'-2)]
The third party applies the block-wise negative-positive transformation to ${ Q}_U^e$. In this step, a transformed pixel value in the $j$th block $B_j$ , $p'$ is 
\begin{equation}
\label{eq:np}
\begin{cases}
p'=p,\ j=\textrm { 0, 2, 4,} \cdots\\
p' = 255 - p,\  j=\textrm { 1, 3, 5,} \cdots
\end{cases}
\end{equation} 
where $p$ is the pixel value of ${ Q}_U^e$ with 8 bpp, and the image with transformed pixel values is referred to as $Q^{NP}$. 

\item[b'-3)] 
The third party calculates an image descriptor from ${ Q}^{NP}$ by using the stored codebook and the stored image descriptors.

\item[b'-4)] 
The third party retrieves EtC images in the database similar to the query image by using the image descriptor without decryption. The retrieved images and the owner's information are returned to the user.

\item[b'-5)]
The user requests the image owner to send key sets for decrypting the EtC images received from the third party.
\end{enumerate}
In the proposed framework, the third party performs the same operations in both system models. 
The transform in steps b-2) and b'-2) enables a mixture of plain and encrypted images to be used.
\section{Experiment}
\begin{figure}[t]
\begin{center}
\begin{tabular}{c}
\begin{minipage}{0.22\hsize}
   \includegraphics[width=18mm]{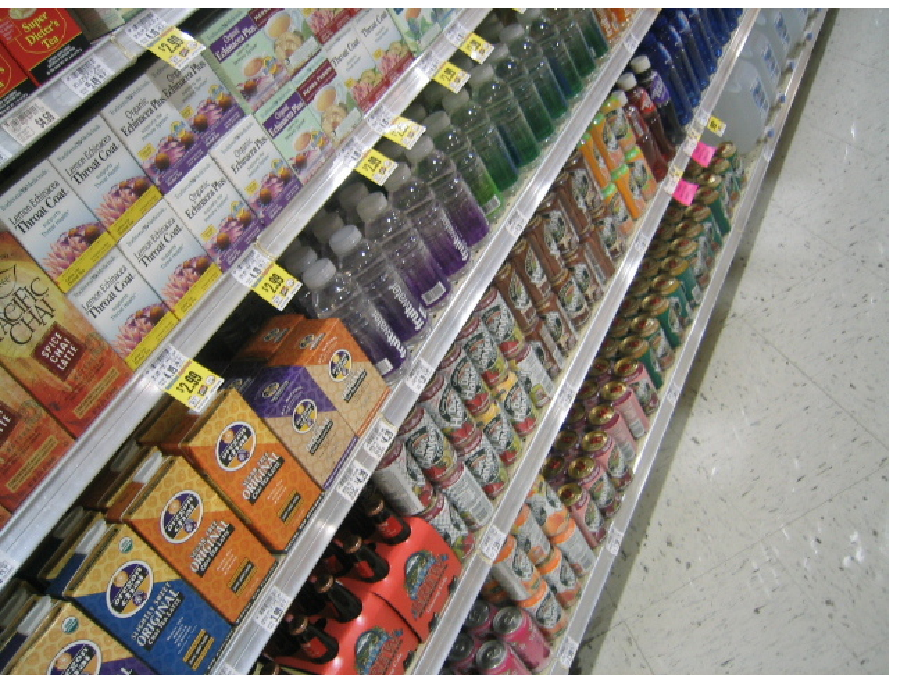}
        %\hspace{16mm}(a) No.00144
 \end{minipage}
\begin{minipage}{0.22\hsize}
   \includegraphics[width=18mm]{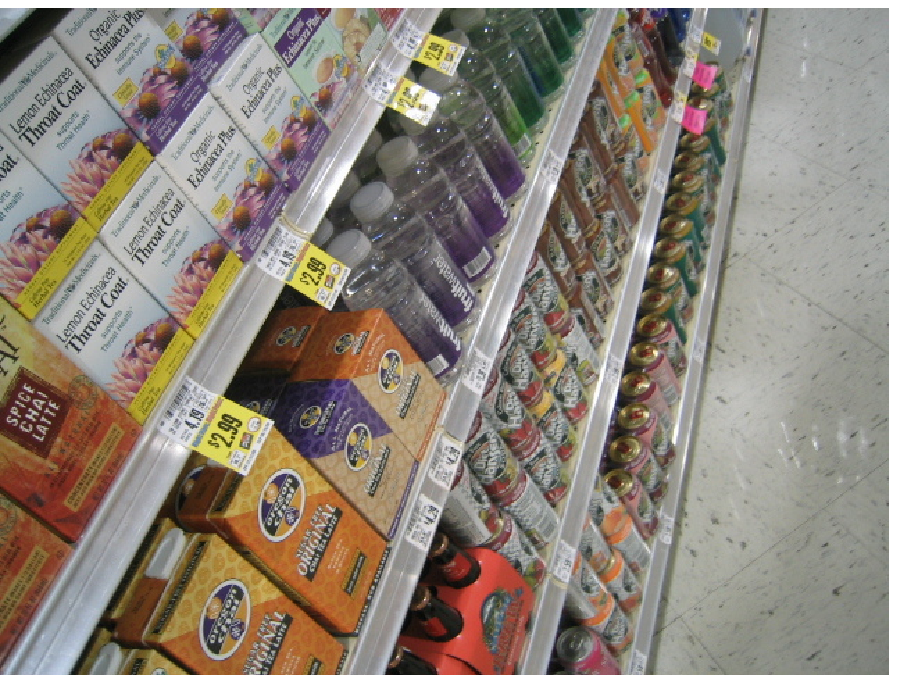}
        %\hspace{16mm}(b) No.00145
 \end{minipage}
 \begin{minipage}{0.22\hsize}
   \includegraphics[width=18mm]{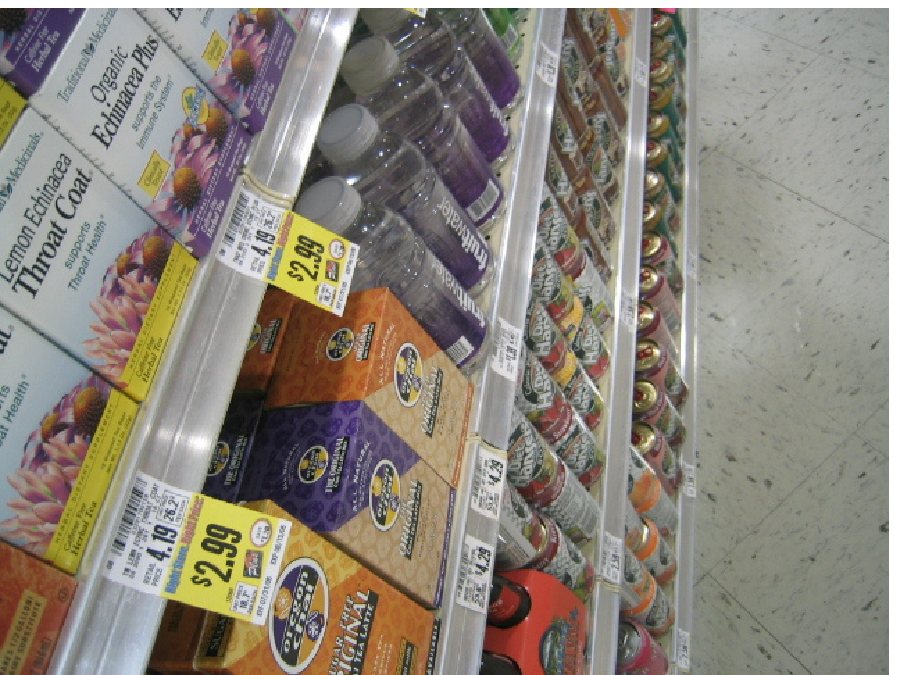}
           %\hspace{16mm}(c) No.00146
 \end{minipage}
 \begin{minipage}{0.22\hsize}
   \includegraphics[width=18mm]{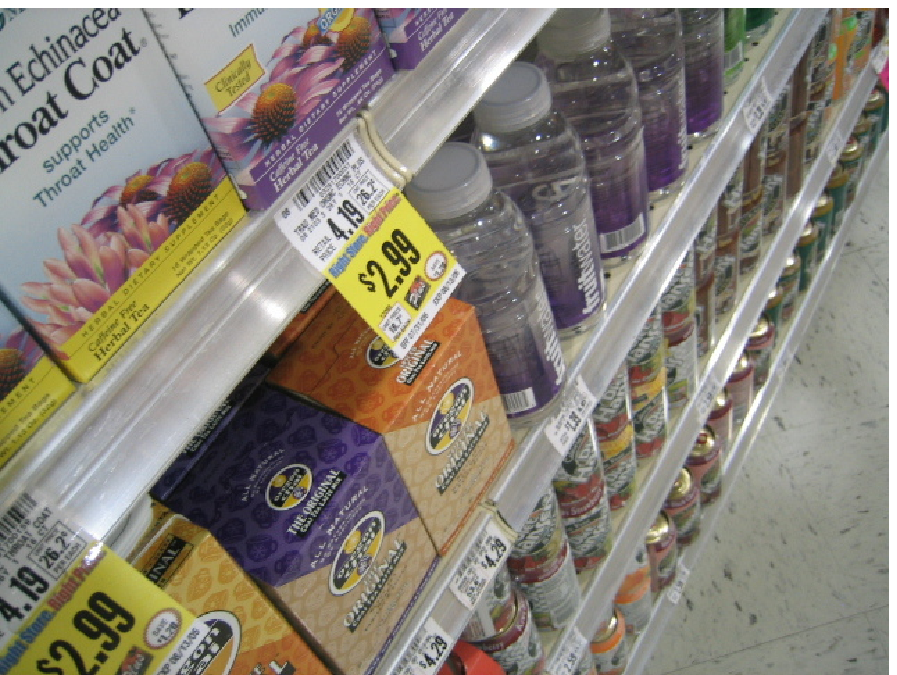}
              %\hspace{16mm}(d) No.00147
 \end{minipage}\\
 (a) Plain images \\
 
 \begin{minipage}{0.22\hsize}
   \includegraphics[width=18mm]{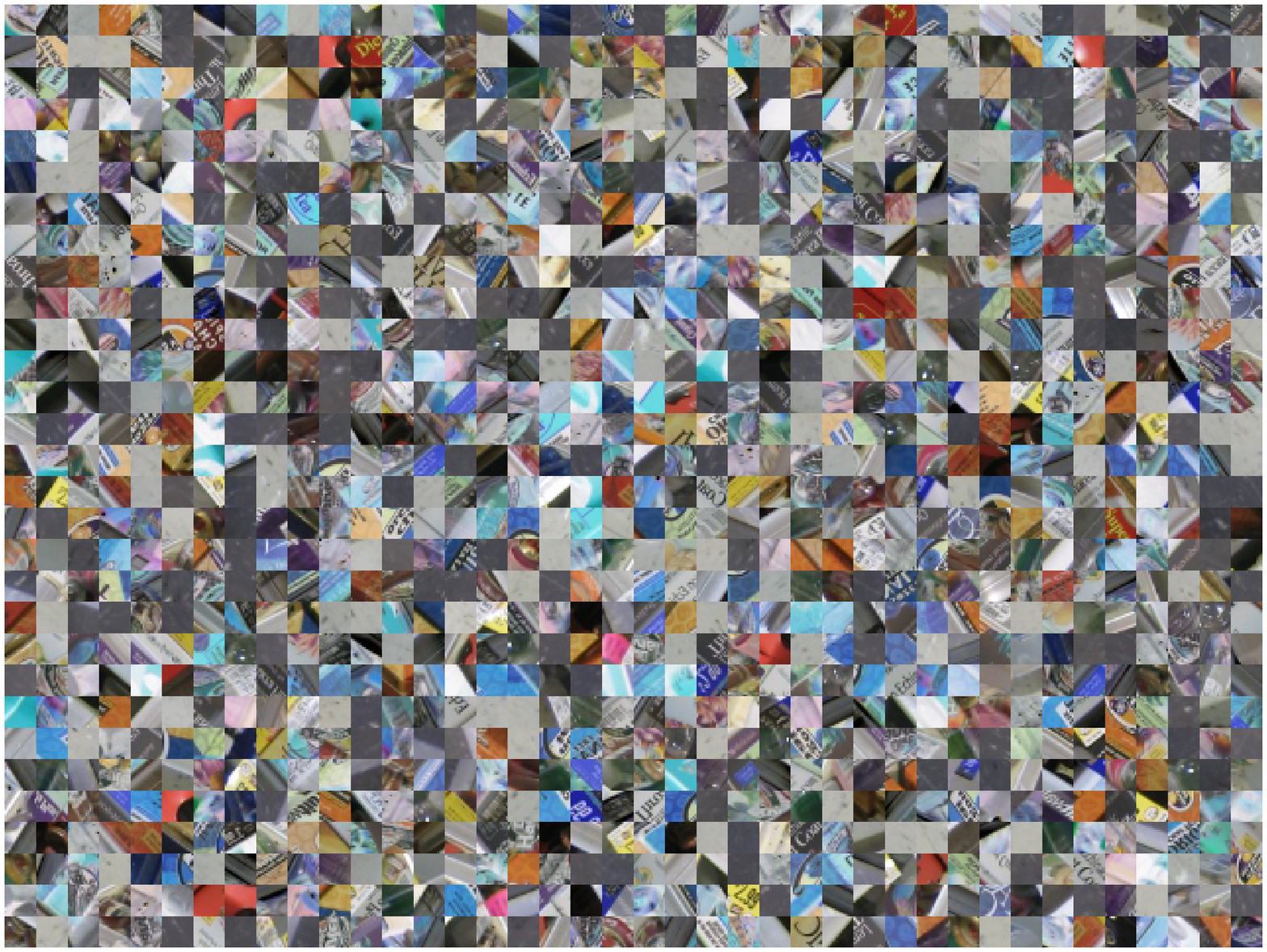}
        %\hspace{16mm}(a) No.00144
 \end{minipage}
\begin{minipage}{0.22\hsize}
   \includegraphics[width=18mm]{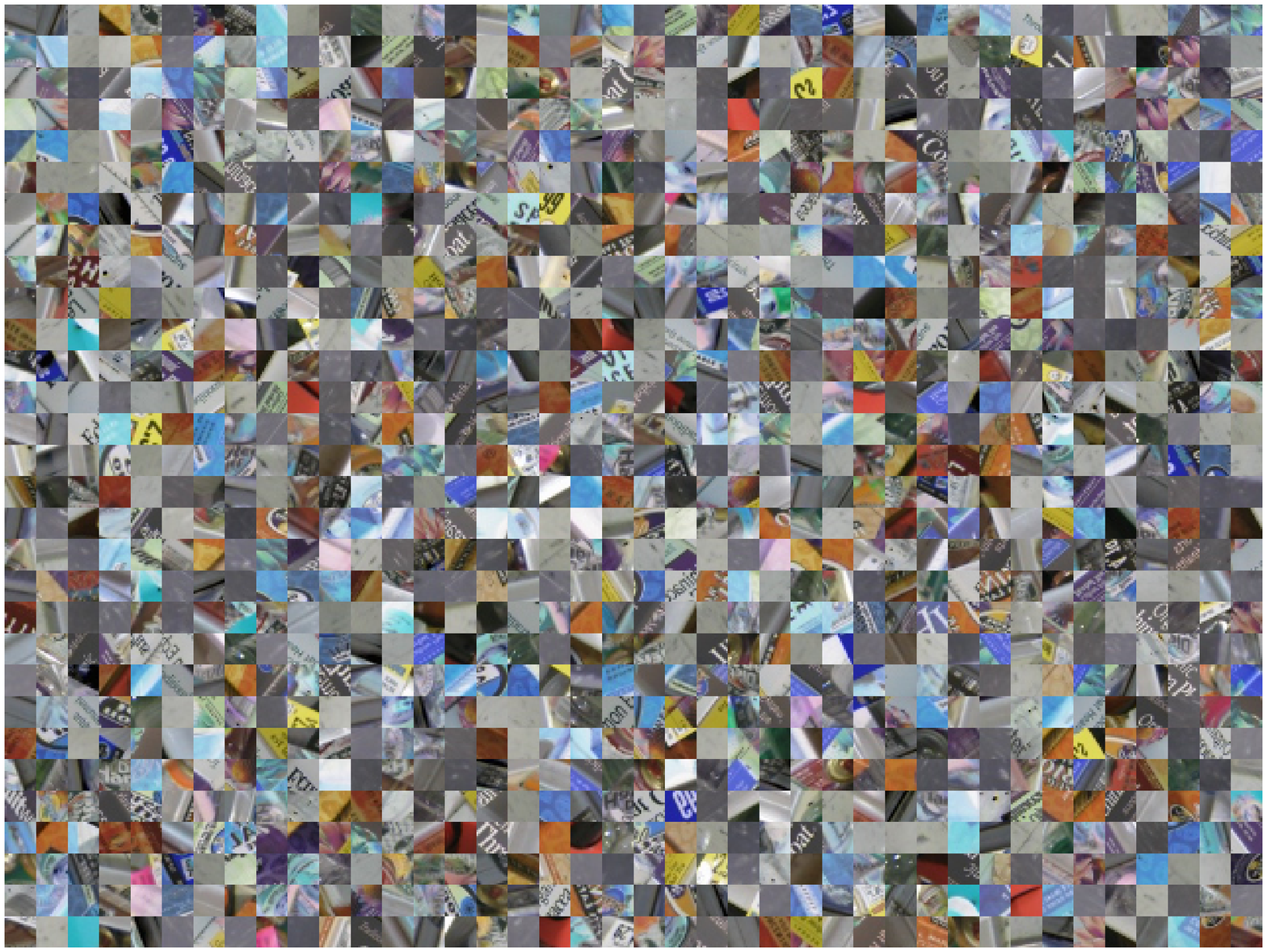}
        %\hspace{16mm}(b) No.00145
 \end{minipage}
 \begin{minipage}{0.22\hsize}
   \includegraphics[width=18mm]{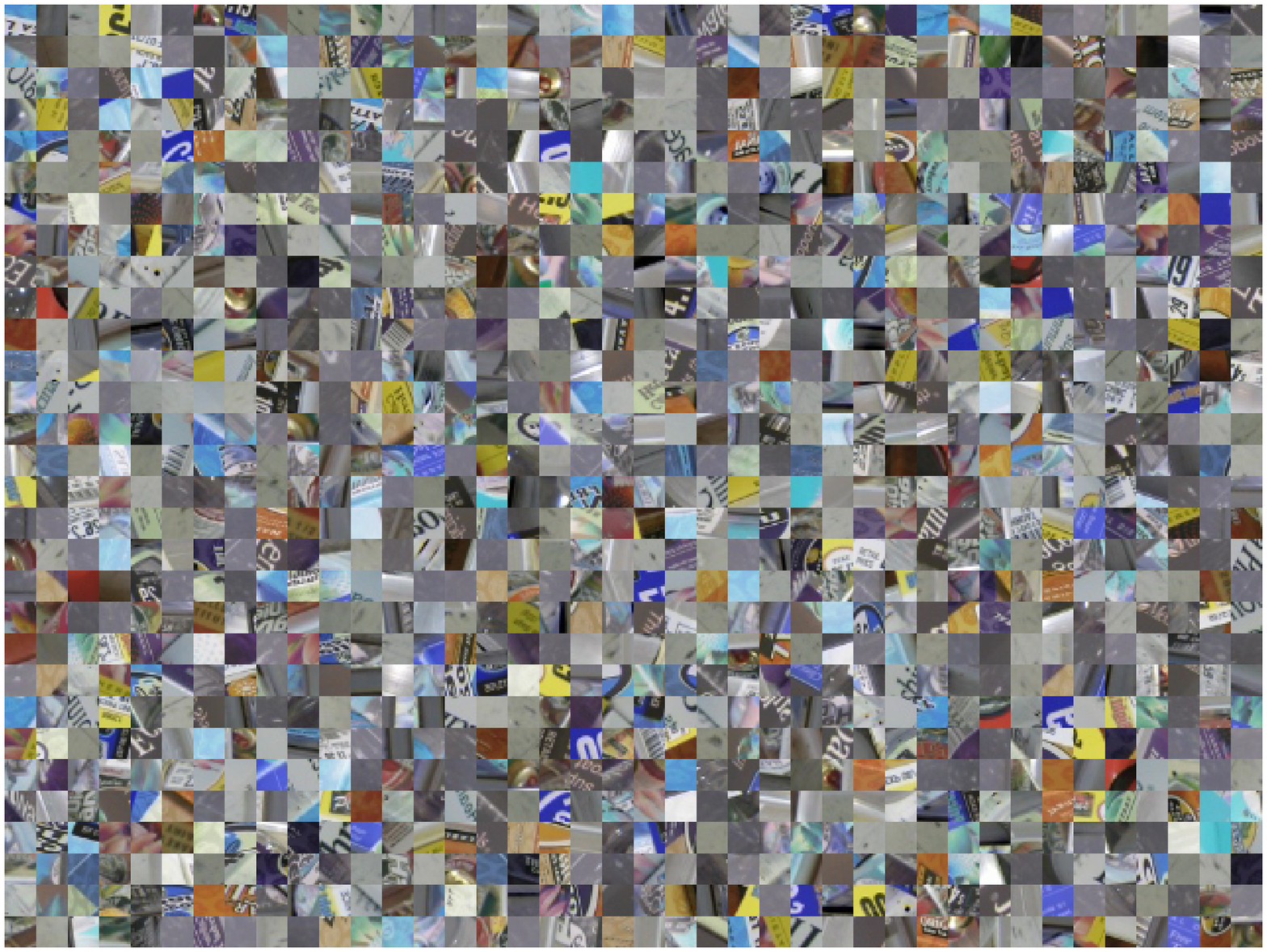}
           %\hspace{16mm}(c) No.00146
 \end{minipage}
 \begin{minipage}{0.22\hsize}
   \includegraphics[width=18mm]{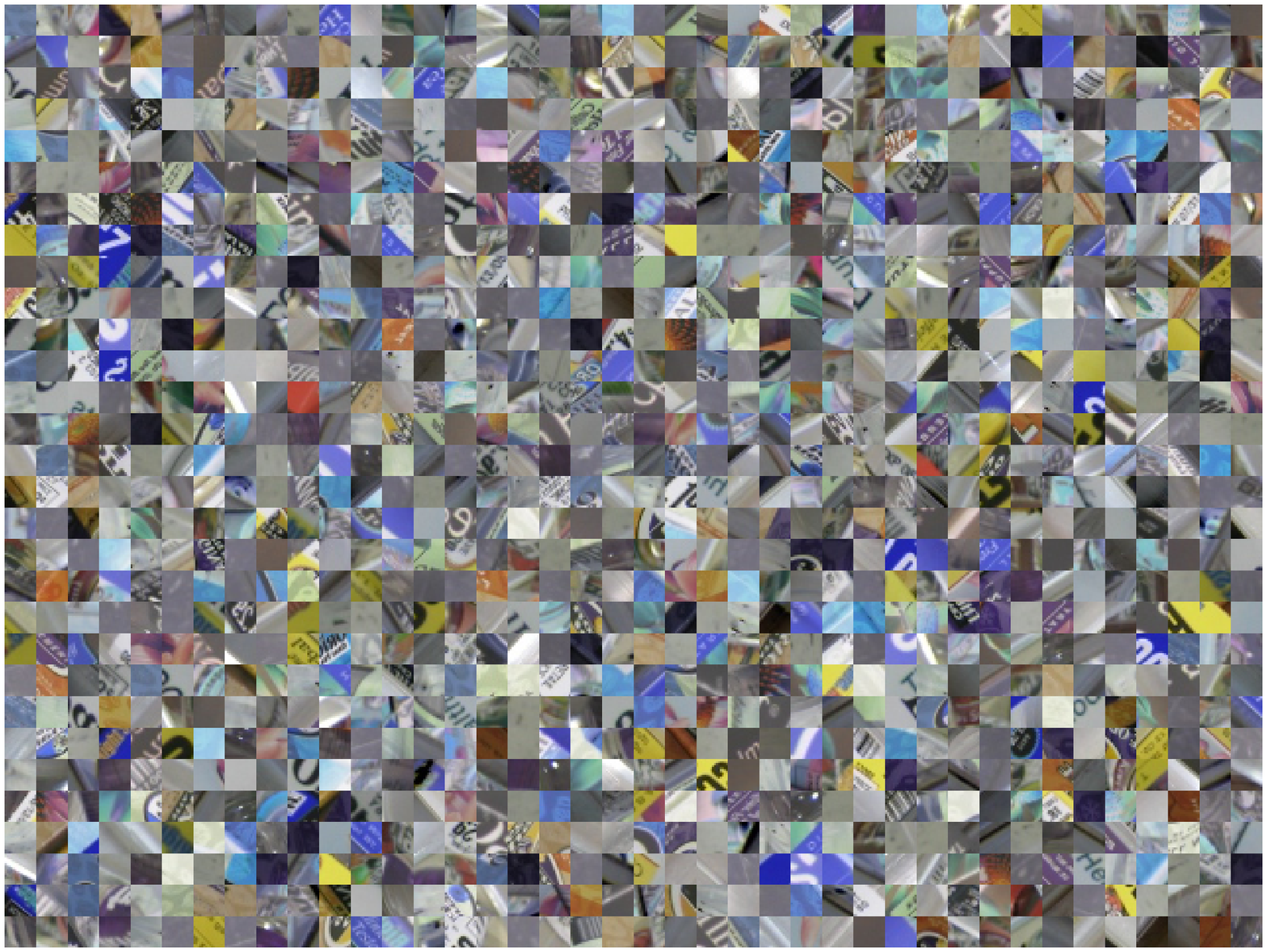}
              %\hspace{16mm}(d) No.00147
 \end{minipage}\\
(b) Corresponding EtC images \\
\end{tabular}
\caption{Image examples in group (UKbench dataset) \label{fig:exuk}}
 \end{center}
\end{figure} 

\begin{table}[t]
\caption{Abbreviated names of relation between stored and query images }
\label{tab:noteSim}
\centering
\scalebox{.95}{
\begin{tabular}{|c|c|c|c|c|c|c|c|c|}\hline
Notation & \shortstack{Stored images} & \shortstack{Query images}\\\hline
``plain images vs plain images" & plain & plain \\\hline
``EtC images vs plain images" & EtC & plain \\\hline
``EtC images vs plain images with NP" & EtC & plain with NP \\\hline
``EtC images vs EtC images" & EtC & EtC \\\hline
\end{tabular}
}
\end{table}
\begin{figure}[t!]
\centering
\includegraphics[width=85mm]{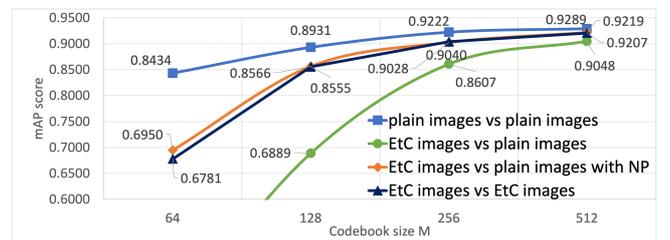}
\caption{Retrieval performance \label{fig:resUK2a}}
\end{figure}
\noindent In this experiment, the retrieval performance of the proposed scheme under the mixed use of EtC and plain images was evaluated in terms of mean average precision (mAP) score. 
For the evaluation with this dataset, the first 1,000 images with a size of 640$\times$480 in UKbench dataset were selected, where they were classified into 250 groups, and each group had 4 images (see Fig. \ref{fig:exuk})\cite{uk}. 
All 1,000 images were used as images stored in the database of the third party, and 250 images were selected as query images from the first images in each group.
%\noindent We evaluated the retrieval performance of the proposed scheme under the mixed use of EtC and plain images in terms of mean average precision (mAP) score. 
%For the evaluation with this dataset, the first 1,000 images with a size of 640$\times$480 in UKBench dataset were selected, where they were classified into 250 groups, and each group had 4 images (see Fig. \ref{fig:exuk})\cite{uk}. 
%In this experiment, 1,000 images were stored in the database of the third party, and the first images in each group, i.e., 250 images, were selected as query images.

\subsection{Effect of image encryption}
\noindent Figure \ref{fig:resUK2a} shows the results of the experiment under four conditions  shown in Tab. \ref{tab:noteSim}. 
From the figure, mAP scores for ``EtC images vs plain images," which is the conventional method\cite{PPIR2}, were heavily degraded,  compared with the other conditions.
In contrast, mAP scores for ``EtC images vs plain images with NP," which is the proposed method, were the same as those for ``EtC images vs EtC images," where plain images with NP indicate images transformed with the negative-positive transform $Q^{NP}$. 
In addition, mAP scores for ``EtC images vs EtC images" under a value of $M$ were almost the same as those for ``plain images vs plain images" under a value of $\frac{M}{2}$.
From these results, by choosing a proper codebook size, the proposed scheme allows us to achieve almost the same retrieval accuracy as that for plain images, even under the mixed use of encrypted and plain images. 

\subsection{Comparison with conventional methods}
\noindent To confirm whether the proposed scheme has sufficient retrieval performance, the performance of the scheme was compared with those of conventional CBIR methods for plain images.
For the comparison with the scheme, five image descriptors were used: scalable color descriptor (SCD) \cite{scd}, color and edge directivity descriptor (CEDD) \cite{cedd}, SURF \cite{surf}, weighted SIMPLE descriptor with random sampling, and weighted SIMPLE descriptor with SURF detector. 
In the case of using the SURF descriptor for retrieval, the bag-of-visual words model and weighting term frequencies were used. 
To generate weighted SIMPLE descriptors, SCD was selected as the type of patch descriptor.

Table \ref{tab:resUKorg} shows the retrieval performances for plain images. 
It was confirmed that the proposed scheme had a higher retrieval performance than the conventional CBIR methods using plain images. 
Therefore, the proposed scheme enables a high retrieval performance to be achieved even under the mixed use of plain and EtC images. 
We also confirmed that the performance for compressing EtC images with JPEG had the same trend as one for compressing plain images with JPEG.

\begin{table}[t!]
\caption{Comparison with conventional CBIR methods using plain images \label{tab:resUKorg}}
\centering
\scalebox{1}{
 \renewcommand{\arraystretch}{1.2}
\begin{tabular}[b]{|c|c|c|c|c|}\hline
\multicolumn{2}{|c|}{Descriptor}&$M=$&mAP score\\\hline
\multicolumn{2}{|c|}{SCD \cite{scd} (plain)}& - & 0.9179\\\hline
\multicolumn{2}{|c|}{CEDD \cite{cedd} (plain)}& - & 0.8806\\\hline
\multicolumn{2}{|c|}{SURF \cite{surf}}&256&0.8304 \\ \cline{3-4}
\multicolumn{2}{|c|}{(plain)}&512&0.8355\\\hline
\multicolumn{2}{|c|}{Weighted SIMPLE }&256&0.9110\\\cline{3-4}
\multicolumn{2}{|c|}{with random sampling (plain) \cite{SIMPLE}}&512&0.9262\\\hline
\multicolumn{2}{|c|}{Weighted SIMPLE}&256&0.8949\\\cline{3-4}
\multicolumn{2}{|c|}{with SURF detector (plain) \cite{SIMPLE}}&512&0.9109\\\hline \hline
\multicolumn{2}{|c|}{E-SIMPLE (proposed)}&256&0.9098\\\cline{3-4}
\multicolumn{2}{|c|}{(``EtC images vs plain images with NP")}&512&0.9219\\\hline
\end{tabular}
}
\end{table}

%%%%%%%%%%%%%%%%%%%%%%%%%%%%%%%%%%%%%%%%%%%%%%%%%%
\section{Conclusion}
\noindent A privacy-preserving content-based image-retrieval scheme allowing a mixture of plain and encrypted images to be used was proposed in this paper. 
In the proposed scheme, EtC images are used as visually protected images, and extended SIMPLE descriptors are applied to EtC images. 
As a result, the scheme enables us to retrieve EtC images even under this mixed use. 
The result of the experiment showed that the image retrieval performance between EtC and plain images was almost the same between EtC images.
%%%%%%%%%%%%%%%%%%%%%%%%%%%%%%%%%%%%%%%%%%%%%%%%%%
\bibliographystyle{IEEEbib}
\bibliography{ref}
\end{document}